\documentclass[aps,secnumarabic,twocolumn,superscriptaddress]{revtex4}

\usepackage{amssymb}
\usepackage{graphicx}

\begin{document}

\title{Coupling molecular spin states by photon-assisted tunneling}

\author{L. R. Schreiber}
\email[]{l.r.schreiber@tudelft.nl}
\author{F. R. Braakman}
\affiliation{Kavli Institute of Nanoscience, TU Delft, 2600 GA Delft, The Netherlands}
\author{T. Meunier}
\affiliation{Kavli Institute of Nanoscience, TU Delft, 2600 GA Delft, The Netherlands}
\affiliation{Institut N{\'e}el, CNRS and Universit{\'e} Joseph Fourier, Grenoble, France}
\author{V. Calado}
\affiliation{Kavli Institute of Nanoscience, TU Delft, 2600 GA Delft, The Netherlands}
\author{J. Danon}
\affiliation{Dahlem Center for Complex Quantum Systems, Freie Universit\"at Berlin, Germany}
\author{J. M. Taylor}
\affiliation{National Institute of Standards and Technology, University of Maryland, USA}
\author{W. Wegscheider}
\affiliation{Institute for Experimental and Applied Physics, University of Regensburg, Germany}
\affiliation{Solid State Physics Laboratory, ETH Zurich, Switzerland}
\author{L. M. K. Vandersypen}
\affiliation{Kavli Institute of Nanoscience, TU Delft, 2600 GA Delft, The Netherlands}

\date{\today}

\def \ni {{\vskip 0.5 cm}}
\def \pz {{\noindent}}
\def \beq {\begin{equation}}
\def \eeq {\end{equation}}
\def \ba {\begin{eqnarray}}
\def \ea {\end{eqnarray}}
\newcommand{\upp}{\hspace{-0.2 pt}\uparrow}
\newcommand{\downn}{\hspace{-0.2 pt}\downarrow}
\newcommand{\dt}{\mbox{d}t}
\newcommand{\ketbrad}[1]{|#1\rangle\!\langle #1|}
\newcommand{\mb}[1]{\mbox{\boldmath$#1$}}
\newcommand{\mc}[1]{{\mathcal #1}}
\newcommand{\mean}[1]{\langle#1\rangle}
\newcommand{\Mean}[1]{\left\langle#1\right\rangle}
\def\ket#1{\left| #1\right>}
\def\bra#1{\left< #1\right|}

\begin{abstract}
Artificial molecules containing just one or two electrons provide a powerful platform for studies of orbital and spin quantum dynamics in nanoscale devices. A well-known example of these dynamics is tunneling of electrons between two coupled quantum dots triggered by microwave irradiation. So far, these tunneling processes have been treated as electric dipole-allowed spin-conserving events. Here we report that microwaves can also excite tunneling transitions between states with different spin. In this work, the dominant mechanism responsible for violation of spin conservation is the spin-orbit interaction. These transitions make it possible to perform detailed microwave spectroscopy of the molecular spin states of an artificial hydrogen molecule and open up the possibility of realizing full quantum control of a two spin system via microwave excitation.
\end{abstract}

\maketitle

In recent years, artificial molecules in mesoscopic systems have drawn much attention due to a fundamental interest in their quantum properties and their potential for quantum information applications. Arguably, the most flexible and tunable artificial molecule consists of coupled semiconductor quantum dots that are defined in a 2-dimensional electron gas using a set of patterned electrostatic depletion gates. Electron spins in such quantum dots exhibit coherence times up to $200$~$\mu$s \cite{Bluhm10}, about $10^4 - 10^6$ times longer than the relevant quantum gate operations \cite{Petta05,Koppens06}, making them attractive quantum bit (qubit) systems \cite{Hanson07}.

The molecular orbital structure of these artificial quantum objects can be probed spectroscopically by microwave modulation of the voltage applied to one of the gates that define the dots \cite{Wiel03}. In this way, the delocalized nature of the electronic eigenstates of an artificial hydrogen-like molecule was observed \cite{Oosterkamp98,Petta04}. More recently, electrical microwave excitation was used for spectroscopy of single spins \cite{Nowack07,Laird07,Pioro08} and coherent single-spin control \cite{Nowack07,Pioro08}, via electric dipole spin resonance (EDSR).

Here, we perform microwave spectroscopy \cite{Oosterkamp98,Petta04,Petersson09} on molecular spin states in an artificial hydrogen molecule formed by a double quantum dot (DD) which contains exactly two electrons. In contrast to all previous PAT experiments, we observe not only the usual spin-conserving tunnel transitions, but also transitions between molecular states with different spin quantum numbers. We discuss several possible mechanisms and conclude from our analysis that these transitions become allowed predominantly through spin-orbit (SO) interaction. The possibility to excite spin-flip tunneling transitions lifts existing restrictions in our thinking about quantum control and detection of spins in quantum dots, and allows universal control of spin qubits without gate voltage pulses.

\section{Device and excitation protocol}

\begin{figure}
\includegraphics[width=8.5 cm]{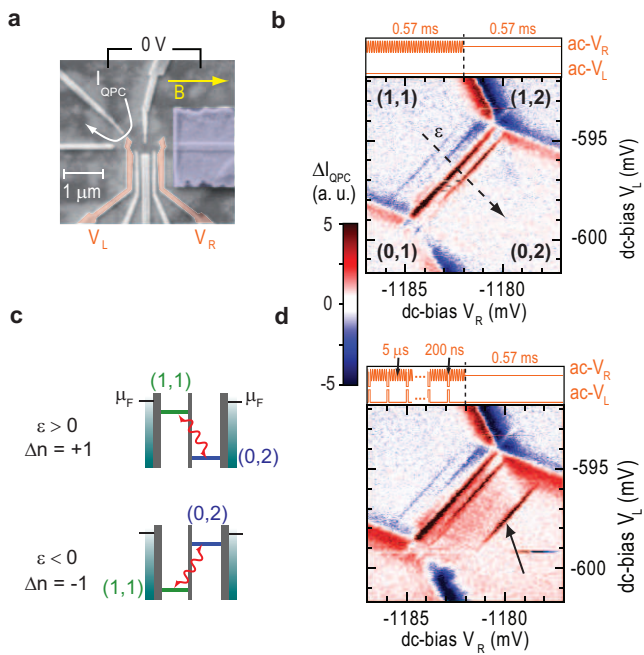}
\caption{\label{fig1}%
\textbf{Photon-assisted tunneling in a 2-electron double quantum dot.} \textbf{a}, Scanning-electron micrograph top view of the double dot gate structure with Co micromagnet (blue). The voltages applied to the left $V_L$ and right $V_R$ side gates (red) control the detuning $\varepsilon$ of the double dot potential. The double dot charge state is read out by means of the current $I_{QPC}$ running through a nearby quantum point contact (white arrow). \textbf{b}, Charge stability diagram around the 2-electron regime at $B=1.5$~T. $(n_L,n_R)$ indicate the absolute numbers of electrons in the left and right dot, respectively. During measurements, 11~GHz microwaves with 880~Hz on-off modulation are applied to the right side gate. The top panel displays schematically one cycle of the ac signal applied to the right side gate. Along the detuning axis (dashed arrow) PAT-lines are observed. \textbf{c}, In the conventional picture of PAT, the first sidebands seen in \textbf{b} should appear when the detuning of the $(0,2)$ and $(1,1)$ states matches the photon energy and interdot tunneling is induced. Further sidebands are then interpreted as multi-photon transitions. $\mu_F$ is the chemical potential of the left and right electron reservoir.
\textbf{d}, Same as in \textbf{b}, but the microwaves are interrupted every 5~$\mu$s by a 200~ns, $P_\varepsilon = 2$~mV detuning pulse applied to the left and right side gates (see the schematic in the top panel). The pulses generate a reference line (black arrow) due to mixing at the $ST^+$ anti-crossing.
}
\end{figure}

Fig. 1a displays a scanning electron micrograph of a sample similar to that used in the experiments. It shows the metal gate pattern that electrostatically defines a DD and a quantum point contact (QPC) within a GaAs/(Al,Ga)As two-dimensional electron gas. An on-chip Co micro-magnet ($\mu$magnet) indicated in blue in Fig. 1a generates an inhomogeneous magnetic field across the DD, which adds to the homogeneous external in-plane magnetic field $B$ (see the appendix A for more sample details), but is not needed for the molecular spin spectroscopy. The sample was mounted in a dilution refrigerator equipped with high-frequency lines. The gate voltages are set so that the DD can be considered as a closed system (the interdot tunneling rates are $10^4$ times larger than the dot-to-lead tunneling rates), and the tilt of the DD potential is tuned by the dc-voltages $V_L$ and $V_R$, applied to the left and right side gates. Working near the turn-on of the first conductance plateau, the current through the QPC, $I_{QPC}$, depends upon the local charge configuration and provides a sensitive meter for the absolute number of electrons $(n_L,n_R)$ in the left and right dot, respectively \cite{Field93,Johnson05}.

First, we excite the DD as indicated in the top panel of Fig. 1b, by adding to $V_R$ continuous-wave microwave excitation at fixed frequency $\nu=11$~GHz. When the photon energy of the microwaves matches the energy splitting between the ground state and a state with a different charge configuration, a new steady-state charge configuration results, which is visible as a change in the QPC current, $\Delta I_{QPC}$. The excitation is on-off modulated at 880~Hz and lock-in detection of $\Delta I_{QPC}$ reveals the microwave-induced change of the charge configuration (see the appendix A for further experimental details). The lower panel of Fig. 1b shows $\Delta I_{QPC}$ as a function of $V_L$ and $V_R$ near the $(1,1)$ to $(0,2)$ boundary of the charge stability diagram. Sharp red (blue) lines indicate microwave-induced tunneling of an electron from the right to the left dot (left to right), labeled as $\Delta n=+1$ ($\Delta n=-1$), respectively (see Fig. 1c). Sidebands can result from multi-photon absorption. At the boundaries with the $(0,1)$ and $(1,2)$ charge states, no energy quantization is observed, since here electrons tunnel to and from the electron-state continuum of the leads. At first sight, the observations in Fig. 1b thus appear to be well explained by the usual spin-conserving PAT processes.

Surprisingly, the position in gate voltage of the resonant lines exhibits a striking dependence on the in-plane magnetic field, $B$. This is clearly seen in Figs. 2a and 2b, which display the measured PAT spectrum along the DD detuning $\varepsilon$ axis (dashed black arrow in Fig. 1b) as a function of $B$ for 20~GHz and 11~GHz excitation, respectively. Since the gate constitutes an open-ended termination of the transmission line, the excitation produces negligible AC magnetic fields at the DD, and is therefore expected to give rise to only electric-dipole allowed spin-conserving transitions, with no $B$ dependence. Furthermore, there is a pronounced asymmetry between the position of the red and blue PAT lines.

In these figures, the detuning axis was calibrated for all magnetic fields by introducing a reference line (see also Fig. 1d, black arrow) that facilitates interpretation of the spectra despite residual orbital effects of the magnetic field. This line was produced by interspersing the microwaves every 5~$\mu$s with 200~ns gate voltage pulses along the detuning axis (see top panel in Fig. 1d), leading to singlet-triplet mixing as described in Ref. \cite{Petta05}. The short gate voltage pulses do not noticeably alter the position of the PAT lines (compare Figs. 1b and 1d). The reference peaks visible at around $\varepsilon=200$~$\mu$eV in Figs. 2a and 2b were aligned by shifting all data points at a given $B$ by the same amount in detuning (see the appendix for the full details of this post-processing step).

\section{Interpretation of the photon-assisted tunneling spectra}

The complexity of the PAT spectra shown in Fig. 2a,b can be understood in detail if we allow for non-spin conserving transitions. The two diagrams in Fig. 2c show the energies of all relevant DD-states (four $(1,1)$-states and one singlet $S(0,2)$-state) as a function of $\varepsilon$ for two different (fixed) magnetic fields, i.e. the spectrum of the DD along the two horizontal dotted lines in Fig. 2b \cite{Koppens05}. Note that the only difference between the two diagrams is the splitting between the three triplet $T(1,1)$-states.

\begin{figure*}
\includegraphics[width=12.0cm]{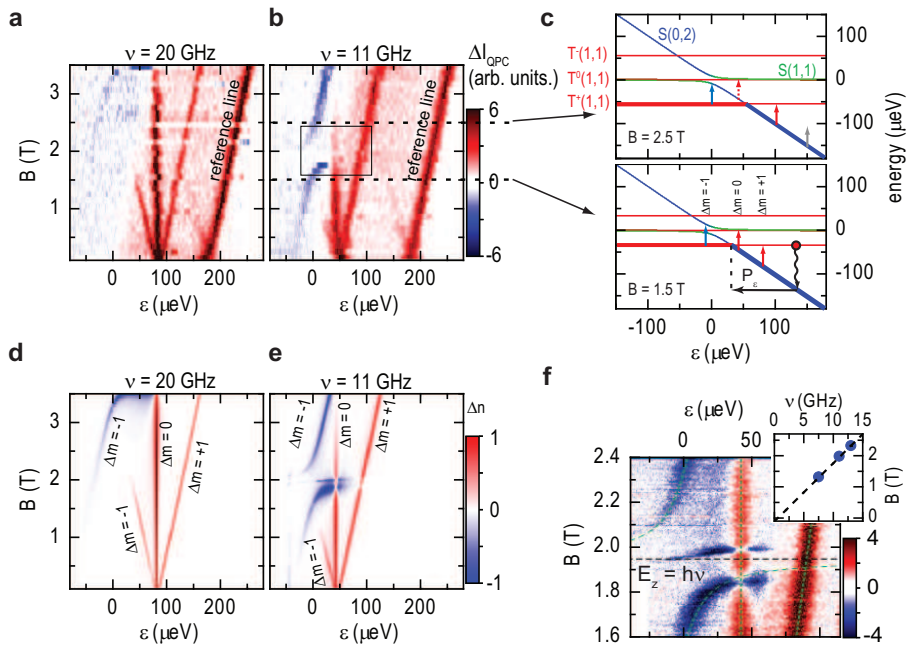}
\caption{\label{fig2}%
\textbf{Photon-assisted-tunneling spectra and simulations.} \textbf{a,b}, Microwave induced change of the QPC current $\Delta I_{QPC}$ as a function of the double dot detuning $\varepsilon$ and the external magnetic field $B$ for 20~GHz (\textbf{a}) and 11~GHz (\textbf{b}) frequency, respectively. singlet-triplet mixing due to 2~mV detuning pulses generates a reference signal that is used to calibrate the detuning axis (see lower panel in Fig. 2f). \textbf{c}, Eigenenergies vs. double dot detuning $\varepsilon$ of the 2-electron spin states in the $(1,1)$ and $(0,2)$ charge regime for two external magnetic fields $B=2.5$~T (upper panel) and $B=1.5$~T (lower panel), respectively. $S(0,2)$, $S(1,1)$ and $T(1,1)$ character of the eigenstates is indicated by blue, green and red color, respectively. The molecular spin ground state is indicated by thick lines. The vertical arrows indicate PAT transitions for a constant microwave frequency involving spin flips. The transition indicated by a dashed arrow is suppressed, because the initial state lies above the ground state. The red circle in the lower panel indicates the detuning position of the reference signal, that is generated by a detuning pulse with amplitude $P_\varepsilon$ to the $ST^+$ anti-crossing. \textbf{d,e}, Simulated PAT spectra for 20~GHz (\textbf{a}) and 11~GHz (\textbf{b}) frequency, respectively. The color indicates the change of the population of the steady-state charge state $\Delta n$ as would be observed in an on-off lock-in detection. A finite temperature of $100$~mK and spontaneous relaxation via the phonon bath are taken into account. \textbf{f}, $I_{QPC}$ scanned with higher resolution in the anti-crossing region (black rectangle in Fig. 2b) at 11~GHz excitation. The green dashed lines indicate the expected detuning positions of the PAT transitions. The horizontal black dashed line indicates the magnetic field, at which the electron spin resonance condition is fulfilled $E_z=g \mu_B (B+b_0)=h \nu$. The graph is concatenated from two scans that overlap at 1.9~T. The inset displays three magnetic fields (blue dots), at which the center of the horizontal blue triplet resonance line is observed, as a function of the microwave frequency $\nu$. The dashed black line gives the expected position of the triplet resonance.}
\end{figure*}

First we explain the resonances observed along the upper dotted line in Fig. 2b ($B = 2.5$ T). In the corresponding (upper) diagram in Fig. 2c, we plot the ground state energy for all $\varepsilon$ with a thick line. If the microwave excitation is off-resonance with all transitions, the system will be in this ground state. For instance, at $\varepsilon = 150~\mu$eV, there is no state available 11~GHz above the $S(0,2)$ ground state (gray arrow) and the system stays in $S(0,2)$. However, when decreasing the detuning, at some point $T^+(1,1)$ becomes energetically accessible (red arrow) and, since we allow for non-spin-conserving transitions, is populated due to the microwave excitation. For this PAT transition, the spin projection on the quantization axis is changed by $\Delta m = +1$. The resulting change of steady-state charge population (increased population of $(1,1)$, or $\Delta I_{QPC} > 0$) is detected by the QPC and yields the red peak in Fig. 2b. Decreasing $\varepsilon$ further, there are two more resonances detectable: (i) the $S(0,2)$-$S(1,1)$ transition (dotted red arrow, $\Delta m=0$), although the signal will be weakened due to the fact that $S(0,2)$ is not unambiguously the ground state anymore. Note that a transition $S(0,2)$-$T^0(1,1)$ could appear in nearly the same detuning position, as will be discussed below. (ii) the $T^+(1,1)$-$S$ transition (blue arrow), where $S$ stands for the hybridized $S(0,2)-S(1,1)$ singlet, results in a negative (blue, $\Delta I_{QPC} < 0$,$\Delta m=-1$) signal from the charge detector since the ground state is now $(1,1)$ and the excited state is $(0,2)$. We see that this simple analysis explains both the positions and the signs of the resonances observed in the data.

A similar analysis can be made for other magnetic fields. For instance, for the spectrum plotted in the lower diagram of Fig. 2c we find two resonances with $\Delta I_{QPC} > 0$ (red arrows), and one with $\Delta I_{QPC} < 0$ (blue arrow). Note that the `blue' transition now connects the ground state to the other branch of the hybridized $S$ compared to the high magnetic field case. Indeed, the singlet anti-crossing is directly probed, resulting in the two blue curved lines observed in the data around $\varepsilon = 0$ (Fig. 2b). The fading out of the blue signal at low fields can be understood from pumping into the metastable state $S(1,1)$: the microwaves excite the system from $T^+(1,1)$ to $S(0,2)$, from where it relaxes quickly to $S(1,1)$. However, relaxation from $S(1,1)$ back to the ground state is slow due to the small energy difference of this transition and the small phonon density of states at low energies \cite{Johnson05, Meunier07}. This pumping weakens the detector signal, since $S(1,1)$ has the same charge configuration as the ground state.

In order to verify this interpretation, we calculate in Figs. 2d,e the position and intensity of the spectral lines at fixed microwave frequency, based on the energy level diagram of Fig. 2c. In the simulations, all single-photon transitions between the ground state and the excited states are allowed by including a matrix element $|T^{\pm}(1,1) \rangle \langle S(0,2)|$ (see appendix A). The input parameters for the calculation of the resonant positions are the interdot tunnel coupling $t_c$, the absolute electron $g$-factor $|g|$, a magnetic-field contribution $b_0$ from the $\mu$magnet parallel to $B$ as well as a magnetic field gradient $\Delta \vec{B} =(\Delta B_x^\bot,0,\Delta B^\|)$ between the dots (in Fig. 3, we show how $t_c$, $|g|$ and $b_0$ can be extracted from the experimental spectra). The color scale represents the calculated steady-state $\Delta n$ that results from microwave excitation, orbital hybridization and phonon absorption and emission at 100~mK. All the PAT transitions visible in the simulation also appear in the experiment, with excellent agreement in both the position and relative intensity of the spectral lines. Especially, the vanishing signal due to spin pumping is also predicted by the simulations which include phonon relaxation.

When we zoom in on the boxed region of Fig. 2b, we see an additional horizontal blue feature at $B \approx 2$~T (Fig. 2f) that also appears in the calculated spectra of Fig. 2e. This feature is due to a triplet resonance from $T^+(1,1)$ to $T^0(1,1)$ that becomes detectable by relaxation into the meta-stable $S(0,2)$ state. In the detuning range where this line appears, the $S(0,2)$ state lies energetically only slightly above the $T^+(1,1)$ state, so relaxation back to the $T^+(1,1)$ ground state is suppressed, again by the small phonon density of states at low energies (see Fig. S2a of the appendix). The triplet resonance is expected to appear at $E_z= g \mu_B (B+b_0)= h \nu$, where $h$ is Planck's constant and $\mu_B$ the Bohr magneton. The inset of Fig. 2f shows the magnetic fields corresponding to the center of the measured triplet resonance line for three excitation frequencies (see Figs. S2b,c of the appendix for the spectra), which are in good agreement with the expected positions (black dashed lines in Fig. 2f) based on the values $|g|$ and $b_0$ determined in the next section from other features of the PAT spectra. Surprisingly, the measured triplet resonance exhibits a finite slope in the $B(\varepsilon)$ spectra. A longitudinal magnetic field gradient $\Delta B^\|$ gives rise to such a detuning dependence, but the $\Delta B^\|$ required in our simulations to reproduce the observed slope is $\Delta B^\| \gtrsim 80$~mT/50~nm, an order of magnitude larger than the gradient we calculate for the $\mu$magnet. The magnitude of the slope remains a puzzle.

\section{Extracting artificial molecule parameters}
\begin{figure}
\includegraphics[width=8.5cm]{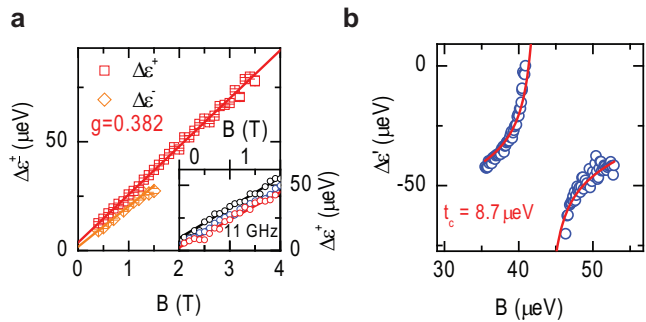}
\caption{\label{fig3}%
\textbf{Analysis of the photon-assisted-tunneling spectra.} \textbf{a}, The detuning difference $\Delta \varepsilon$ between the spin conserving line $\Delta m=0$ and the $\Delta m= \pm 1$ line is plotted as a function of the external magnetic field $B$ at 20~GHz (see Fig 2a), in order to fit the absolute effective electron g-factor $g$ from the slopes of the linear fits (solid lines). \textbf{inset}, $\Delta \varepsilon$ for the blue $\Delta m=+1$ line at 11~GHz for different tunnel couplings. The offset of the curves increases with increasing $t_c$ (red to black circles) while $g$ remains constant. \textbf{b}, $\Delta \varepsilon$ between the $\Delta m=0$ line and the $\Delta m= -1$ $(1,1)$ to $(0,2)$ transition from Fig. 2c. The fit of the anti-crossing (red line) allows for a precise determination of $t_c$.
}
\end{figure}

We now show how $|g|$, $t_c$ and $b_0$, the parameters used for all simulations, can be extracted independently from the experimental spin-flip PAT spectra. For this analysis we only use the relative distance $\Delta \varepsilon$ between PAT lines at fixed magnetic field, in order to be independent from the calibration of the detuning axis by means of the reference line. Fig. 3a shows $\Delta \varepsilon^\pm$ as a function of $B$ using the 20~GHz data. $\Delta \varepsilon^\pm$ is defined as the difference in detuning between the red $\Delta m=\pm 1$ and $\Delta m=0$ PAT lines. For a fixed $\nu$, both $\Delta \varepsilon^\pm$ increase linearly with the Zeeman energy and therefore allow fitting of $|g|$. (Note that for $\Delta \varepsilon^+$ the linearity is only exact for sufficiently large $B$, at which the singlet anti-crossing does not affect the $T^+(1,1)$ energy; see the appendix for a detailed discussion). A least-squares fit to the $\Delta \varepsilon^+$ data gives $|g|=0.382\pm 0.004$ (Fig. 3a). From the linear behavior of $\Delta \varepsilon^+$, we also deduce that there is negligible dynamic nuclear polarization in the experiment.

Knowing $|g|$ precisely, we make use of the blue anti-crossing in Figs. 2b and 2f in order to determine $t_c$ (and $b_0$). Fig. 3b shows the difference in detuning $\Delta \varepsilon^\prime$ between the blue $\Delta m=-1$ and the red $\Delta m=0$ lines in Fig. 2f. Assuming the $\Delta m=0$ line corresponds to the $S(0,2)$-$S(1,1)$ transition (as shown below), then

\begin{equation}\label{eq:Ip}
    \Delta \varepsilon^\prime=\frac{t_c^2-\left(h\nu-g \mu_B (B+b_0)\right)^2}{h\nu - g \mu_B (B+b_0)}-\sqrt{(h\nu)^2-(2t_c)^2} \;,
\end{equation}

\noindent where the first term is the detuning position of the $T^+(1,1)$-$S$ transition and the second the one of the $S(0,2)$-$S(1,1)$ transition. The best fits are obtained with $t_c= 8.7\pm 0.1$~$\mu$eV and $b_0=109\pm 16$~mT. This value for $b_0$ matches very well our simulations of the stray field of the $\mu$magnet at the DD location (see appendix A).

An important question left open so far is whether the red $\Delta m=0$ line involves predominantly transitions from $S(0,2)$ to $S(1,1)$ or to $T^0(1,1)$. The transition to $S(1,1)$ does not require a change in the (total) spin and is thus expected to be excited more strongly than that to $T^0(1,1)$. However, relaxation from $S(1,1)$ back to $S(0,2)$ will be stronger as well, so it is not obvious what steady-state populations will result in either case. Furthermore, given the small energy difference between $S(1,1)$ and $T^0(1,1)$, the two transitions are not resolved in Fig. 2. Fig. 3a helps to answer this question: The observation that $\Delta \varepsilon^{+} > \Delta \varepsilon^{-}$ indicates that the $\Delta m=0$ line originates from the transition to $S(1,1)$ and not to $T^0(1,1)$. For the former we expect $\Delta \varepsilon^{\pm}= E_z \pm J$, with $J(\varepsilon,t_c)=\frac{t_c^2}{\varepsilon}+O(t_c^4)$ the exchange energy, whereas the latter would result in $\Delta \varepsilon^{+} \lesssim \Delta \varepsilon^{-}$ (in both scenario's, $b_0$ causes an additional fixed offset in both $\Delta \varepsilon^{\pm}$, but it does not contribute to their difference). This interpretation is consistent with the increase of $\Delta \varepsilon^{+}$ with larger interdot tunnel coupling, hence larger $J$ (Fig. 3a inset; note that the slopes are not affected). It is further supported by the data in Fig. S3b.

So far only single-photon processes were considered, but at higher microwave power, also multi-photon lines emerge (Fig. 4a), mostly for the $S(0,2)$-$S(1,1)$ transition (green dashed lines in Fig. 4a). Like the single-photon $S(0,2)$-$S(1,1)$ line, their position in detuning is $B$-independent (see appendix for details).

\section{Identification of the spin-flip mechanisms}

Having shown the power of spin-flip PAT for detailed molecular spin spectroscopy, we now discuss the mechanisms responsible for this process as confirmed by our simulations. As a first possibility, the transitions from $S(0,2)$ to the triplet $(1,1)$ states can take place through a virtual process involving $S(1,1)$: the state $S(0,2)$ is coupled to $S(1,1)$ by the interdot tunnel coupling, and an (effective) magnetic field gradient $\Delta \vec{B}=(\Delta B^\bot,0,\Delta B^\|)$ across the DD couples the spin part of all the $(1,1)$ states to each other \cite{Johnson05,Taylor07}. Here $\Delta \vec{B}$ has a contribution from the effective nuclear field and from the $\mu$magnet. The transition matrix element from $S(0,2)$ to $T^\pm(1,1)$ is $\propto t_c \frac{\Delta B^\bot}{B}$, assuming $E_z \gg J$. In the following, we use the $B$-dependence of this process as a fingerprint and focus on the red $\Delta m=+1$ line in Figs. 2a and 2b, as we can follow it over the entire magnetic field range. The intensity of this line is constant in $B$ and even if the microwave amplitude $E$ is varied, we observe no $B$-dependence in the area under this peak (Fig. 4b). Before we conclude that the transition rate is magnetic field independent, we recall that the observed PAT lines reflect the steady-state change in the charge configuration resulting from stimulated photon emission and absorption and spontaneous relaxation. In order to rule out that a field-independent steady state is reached from a field dependence of relaxation and excitation that cancel each other, we verify that the spontaneous relaxation rate is field-independent as well (see appendix). These observations suggest that the coupling mechanism is magnetic field independent and thus virtual processes involving $S(1,1)$ do not give a strong contribution to the transition rates.

\begin{figure}
\includegraphics[width=8.5cm]{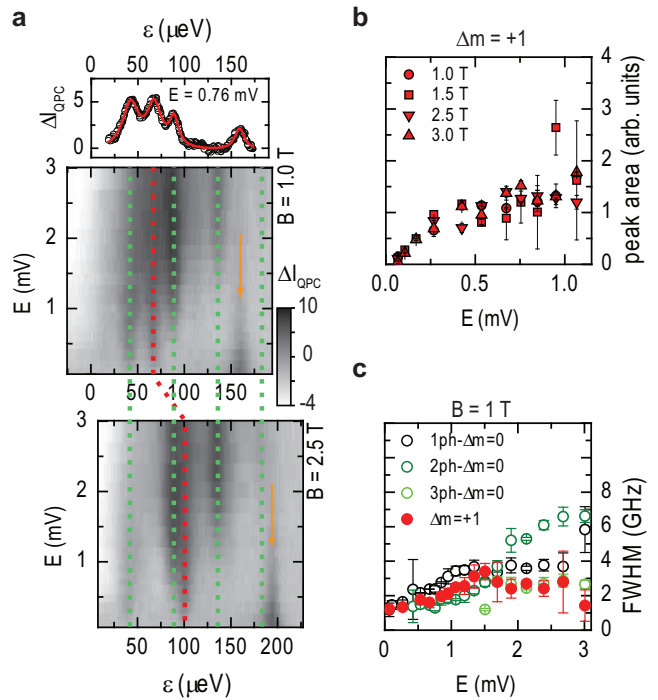}
\caption{\label{fig4}%
\textbf{Power-dependence of the photon-assisted tunneling spectra and spontaneous relaxation.} \textbf{a} $\Delta I_{QPC}$ as a function of the double dot detuning $\varepsilon$ and the microwave amplitude $E$ for 11~GHz and $B=1.0$~T (middle panel) and $B=2.5$~T (lower panel). The excitation scheme from Fig. 1d is employed with reference pulse amplitude $P_\varepsilon=1.5$~mV. The green (red) dashed lines mark the multi-photon $\Delta m=0$ ($\Delta m= +1$) PAT transitions. The reference signal stemming from the pulse is marked by orange arrows. The voltage amplitude $E$ is measured at the end of the coaxial lines at room temperature. The uppermost panel displays a linecut measured at 1~T. The red line is a least-squares fit by a sum of four lorentzian peaks to the data. \textbf{b} The fitted Lorentzian peak area of the $\Delta m= +1$ line from \textbf{a} is plotted as a function of microwave amplitude $E$ and magnetic field $B$. \textbf{c} The fitted Lorentzian linewidth at half maximum (FWHM) of the $\Delta m= +1$ and the lines $\Delta m= 0$ from \textbf{a} is plotted as a function of microwave amplitude $E$ for $B=1$~T. The error bars in \textbf{b},\textbf{c} and are determined from the Lorentzian least-squares fit (see the inset of Fig. 4a).}
\end{figure}

More recently, two mechanisms were considered that provide a direct, $B$-independent matrix element between $S(0,2)$ and the $(1,1)$ triplet states: (i) the hyperfine contact Hamiltonian is of the form $\sum_j \delta(r-r_j) \vec{I}^j \cdot \vec{S}$. Thus, nuclear spins, $\vec{I}_j$, in the barrier regions, where the $S(0,2)$ spatially overlaps with each of the $(1,1)$ triplet states, can flip-flop with the electron spin, $\vec{S}$, simultaneously with charge tunneling \cite{Stopa10}. (ii) The SO Hamiltonian is of the form $p_{x,y} S_{x,y}$ and can directly couple states which differ in both orbital and spin \cite{Nadj10,Danon09}. (When the orbital part of the initial and final state are the same, the SO Hamiltonian does not provide a direct matrix element and the transition rate becomes $B$-dependent \cite{Khaetskii01,Fujisawa02,Golovach06,Nowack07,Golovach08}.) The ratio of the SO mediated rate and the hyperfine mediated rate can be estimated as $\frac{E_0\sqrt{N}}{4A}\frac{d}{\lambda_{SO}}$, which is a few thousand in the experiment (see the appendix). Here $E_0$ is the single-dot level spacing, $N$ the number of nuclei in contact with one dot, $A$ the hyperfine coupling strength, $d$ the interdot distance and $\lambda_{SO}$ the spin-orbit length. We therefore believe that SO interaction is the dominant spin-flip mechanism for the observed PAT transitions. The presence of a magnetic field independent matrix element between $S(0,2)$ and the $(1,1)$ triplets is confirmed by the observation that the intensity of the reference signal shows no field dependence.

Finally, we extract from Fig. 4a the fitted linewidth as a function of driving power (Fig. 4c). For small $E$, we find a width $\gtrsim 1$~GHz, similar to that observed earlier for spin-conserving PAT processes \cite{Oosterkamp98}. For stronger driving, both the $\Delta m=0$ and the $\Delta m=+1$ lines are further broadened, up to $E\approx 1.5$~mV. If these lines were power broadened, their width would imply transition rates in excess of 1~GHz. However, we do not believe that this is the case, since in measurements with short microwave bursts the populations saturated only on a long (10~$\mu$s) timescale (data not shown). Presumably charge or gate voltage noise is responsible for the broadening instead.

We have shown that in our DD system, all $(1,1)$ spin states have at least weakly allowed electric dipole transitions to $S(0,2)$. In materials with high SO interaction like InAs, the effect of the non-spin conserving PAT will be even stronger. In materials with weak SO interaction, a strong gradient magnetic field can be used to facilitate spin-flip tunneling transitions. In all cases, this opens the possibility of realizing full quantum control of the $(1,1)$ spin space via off-resonant (microwave) Raman transitions through the excited $S(0,2)$ state, which enables a variety of new approaches to manipulating and even defining qubits in DDs. Furthermore, such control enables new measurement techniques that do not rely on Pauli spin blockade \cite{Ono02}. An example is a measurement that distinguishes parallel from anti-parallel spins while acting non-destructively on the $S(1,1)-T^0(1,1)$ subspace, by coupling resonantly the $T^+(1,1)$ and $T^-(1,1)$ states to $S(0,2)$ followed by charge readout. This constitutes a partial Bell measurement and leads to a new method for producing and purifying entangled spin states \cite{Taylor05}.

\appendix
\section{Methods}

\subsection{Sample fabrication}
30~nm thick TiAu gates are fabricated on a 90~nm deep (Al$_{0.3}$,Ga$_{0.7}$)As/GaAs two-dimensional electron gas (2DEG) by means of ebeam lithography. The double dot axis is aligned along the [110] GaAs crystal direction (z-direction), which is parallel to the external magnetic field direction $\vec{B}$ (Fig. 1a). The 2DEG is Si $\delta$-doped (40~nm away from the hetero-interface), exhibits an electron density of $2.05 \times 10^{11}$~cm$^{-2}$ and a mobility of $2.06 \times 10^{6}$ cm$^{2}$/Vs at 1~K in the dark. The grounded, 275~nm thick, 2~$\mu$m wide and 10~$\mu$m long Co $\mu$magnet is evaporated on top of a 80~nm thick dielectric layer, aligned along $\vec{B}$ (magnetic easy axis) and placed $\approx 400$ nm away from the closest dot center. We calculate \cite{Goldman00} that at the double dot position the $\mu$magnet adds $b_0 \approx 110$~mT to $B_z$ and generates a magnetic field gradient of $\Delta B^\| \approx 6$~mT/50~nm and a transverse gradient of $\Delta B^\bot \approx -6$~mT/50~nm at saturation ($B_z \gtrsim 2$~T).


\subsection{Measurement}
The sample is mounted in an Oxford KelvinOx 300 dilution refrigerator at 30~mK. Left and right side gate voltages, $V_L$ and $V_R$, are set by low-pass filtered dc lines and $\approx 60$~dB attenuated coaxial lines combined with bias-tees with a cutoff frequency of 30~Hz. The pre-amplified current through the quantum-point contact is read out by a lock-in amplifier locked to the 880~Hz on-off modulation of the microwaves. The bias across the double dot is set to 0~$\mu$V. Voltage pulses to the left and right side gates are generated with a Sony Textronix AWG520. The microwaves are generated with a HP83650A and combined with the pulses to the right side gate. Microwave bursts and detuning pulses are synchronized to ensure that the microwave excitation is switched off during the detuning pulses that generate the reference signal (see Fig. 1d).

\subsection{Simulation}

The Hamiltonian describing the two-spin system near the $(1,1)$-$(0,2)$ transition is taken to be a five-state system, with four $(1,1)$ spin states and a $(0,2)$ spin singlet \cite{Taylor07} in the presence of an external magnetic field $B$ and a magnetic field gradient $\Delta \vec{B}$, which includes both the quasi-static nuclear field and the field from the $\mu$magnet.  This is given by $H = g \mu_B P_{11} (\vec{S}_1 \cdot (\vec{B} + \Delta \vec{B}) + \vec{S}_2 \cdot (\vec{B} - \Delta \vec{B})) P_{11} - \varepsilon \ketbrad{S(0,2)} + H_{t}$, were $P_{11}$ is the projector onto the $(1,1)$ subspace, $\varepsilon$ is the detuning due to the difference in gate potentials from the left and right gates and the tunnel coupling $H_t=t_c (\ket{S(1,1)} \bra{S(0,2)}) + t_{SO} (\ket{T^+(1,1)} \bra{S(0,2)}) + t_{SO} (\ket{T^-(1,1)} \bra{S(0,2)}) + \rm{h.c.}$ with $t_c$ the spin-conserving
tunnel coupling and $t_{SO}$ the spin-orbit coupling set to $5$~\% of $t_c$.

To find the signal we expect theoretically from the experiment, we add a weak, rapidly oscillating term to the Hamiltonian: $\varepsilon \rightarrow \varepsilon_0 + \Omega \cos(\nu t)$.  We diagonalize $H$ with $\Omega = 0$, then make a rotating frame transformation in which levels are grouped into bands $n$ (defined by a projector $P_n$) where the states in a band $n$ are much closer in energy than $\hbar \nu$, while the energy difference between states in band $n$ and $n+1$ are within 2/3rds of $h\nu$.  Each band rotates at a rate $n \nu$, and we can then make a rotating wave approximation, keeping terms due to $\delta$ that couple adjacent bands, i.e., our perturbation in the rotating frame and rotating wave approximation is $V = \Omega \sum_n P_n \ketbrad{(0,2)S} P_{n+1} + {\rm h.c.}$. Next, we add dissipation and dephasing by including relaxation due to coupling of the electron charge to piezoelectric phonons in a two-orbital (Heitler-London-like) model.

\setcounter{figure}{0}
\renewcommand{\figurename}{Fig. S}

\section{Calibration of the detuning axis}

\begin{figure}
\includegraphics[width=8.5cm]{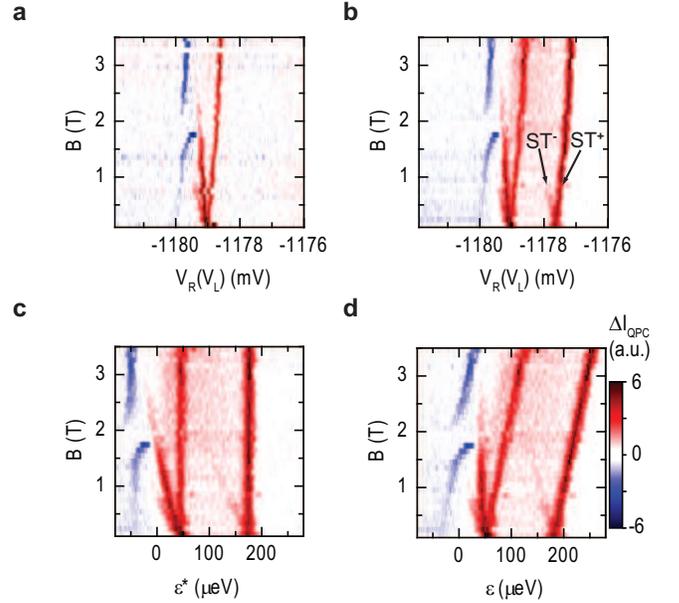}
\caption{\label{fig6}%
\textbf{Calibration of the detuning axis for the PAT spectra.} \textbf{a}, PAT spectrum as measured without pulses (see inset of Fig. 1b) with a microwave frequency of 11~GHz. The voltage applied to the right side gate $V_R$ and left side gate $V_L$ is swept simultaneously to follow the detuning axis indicated in Fig 1b. \textbf{b}, same PAT spectrum, but with 200~ns pulses that intersperse the microwave bursts every 5~$\mu$s (see top panel of Fig. 1d). A red reference line appears at positive DD detuning, when the detuning pulse reaches exactly the $ST^+$ anti-crossing. Also a weak line due to mixing at the $ST^-$ anti-crossing is observed. Both lines are marked by arrows. The more red background is due to $ST^0$ mixing. \textbf{c}, PAT spectrum with calibrated detuning axis. The horizontal lines of the spectrum are shifted so that the reference pulse appears at the detuning position, which equals the pulse amplitude $P_\varepsilon$ converted into energy (see text) along the detuning axis. Here $\varepsilon^*=0$ is equal to the $ST^+$ anti-crossing for every magnetic field. \textbf{d}, To convert the $\varepsilon^*$ scale of the detuning axis into the $\varepsilon$ scale, where $\varepsilon=0$ equals the $S(1,1)$-$S(0,2)$ anti-crossing, the spectrum is sheared by the Zeeman-energy using the electron g-factor that is independently determined as explained in the main article. At low magnetic fields this transformation of scales is wrong and gives rise to a curvature of the spectrum, which is a function of the tunnel coupling.}
\end{figure}

The photon-assisted tunneling (PAT) spectra in Figs. 2a and 2b of the main article are measured both with a high energy resolution along the double dot (DD) detuning axis and over a wide external magnetic field range. In the experiment, we observe a monotonous, reproducible drift of the stable charge regions predominantly along the right side gate voltage as we change the magnetic field. Changing the voltages applied to the left and to the right side gate accordingly, we partially compensate for this drift. We then record an 11~GHz PAT spectrum as displayed in Fig. S1a. In order to precisely calibrate the detuning axis for all $B$, a reference signal is generated together with the PAT spectrum by interspersing the microwaves every 5~$\mu$s by a 200~ns detuning pulse with amplitude $P_\varepsilon$ towards negative detuning (Fig. S1b). 200~ns are found to be sufficient to mix the $S(0,2)$ state entirely with the $T^+(1,1)$ state at their anti-crossing $\varepsilon_{ST+}$, so that a Pauli spin blocked $T^+(1,1)$ signal is observed at a detuning $\varepsilon=\varepsilon_{ST+}+P_\varepsilon$. The magnitude of $P_\varepsilon$ is chosen such that the reference signal appears at a detuning position far away from the PAT signal. The pulses do not alter the detuning position of the PAT resonances. In addition to $ST^+$ mixing, mixing of the $S$ with $T^0(1,1)$ and $T^-(1,1)$ is observed due to the pulsing. The former gives rise to a positive $\Delta I_{QPC}$ background on the left of the reference signal in Fig. S1b. The latter generates a weak second reference line that overlaps with the $ST^+$ reference line for $B \rightarrow 0$~T, but shifts towards negative detuning as $B$ is increased.

In a post-processing step, we separately fit the position of the $ST^+$ peaks for all $B$ and shift every row of the spectrum, such that the peaks are vertically aligned at $\varepsilon^*=P_\varepsilon$ as shown in Fig. S1c. Thus, $\varepsilon^*=0$ is at the $ST^+$ anti-crossing, by definition. All data points at a given $B$ are shifted by the same amount in detuning. This $\varepsilon^*$-detuning axis is well-defined but `moves' with respect to the $\varepsilon$-detuning axis as a function of $B$, since $\varepsilon^*=\varepsilon-\varepsilon_{ST+}(B)$. The lever arm for the voltage to energy conversion is read from the voltage distance of the second and third $\Delta m=0$ PAT line at $\nu=11$~GHz (see Fig. 4a), which equals the photon energy $h \nu$ \cite{Petta04}. For the analysis of all spectra we used the same lever arm.

For better readability of the PAT spectra, we finally convert the $\varepsilon^*$-scale to the $\varepsilon$-scale found in literature, for which $\varepsilon=0$ is defined by the $S(1,1)$ to $S(0,2)$ anti-crossing. To do so, we additionally shift all data points at a given $B$ by $|g| \mu_B |B|$ towards positive detuning (Fig. S1d). However, since $\varepsilon_{ST+}=|g \mu_B B|$ holds true only for $|g \mu_B B|\gg t_c$, where $t_c$ is the interdot tunnel coupling, the detuning axis conversion fails for low magnetic fields. As a result the PAT resonance-lines bend towards positive $\varepsilon$ for $B \rightarrow 0$~T in Fig. S1d and in the spectra shown in the main article (Fig. 2a,b).

\section{Triplet spin resonance}

\begin{figure}
\includegraphics[width=8.5cm]{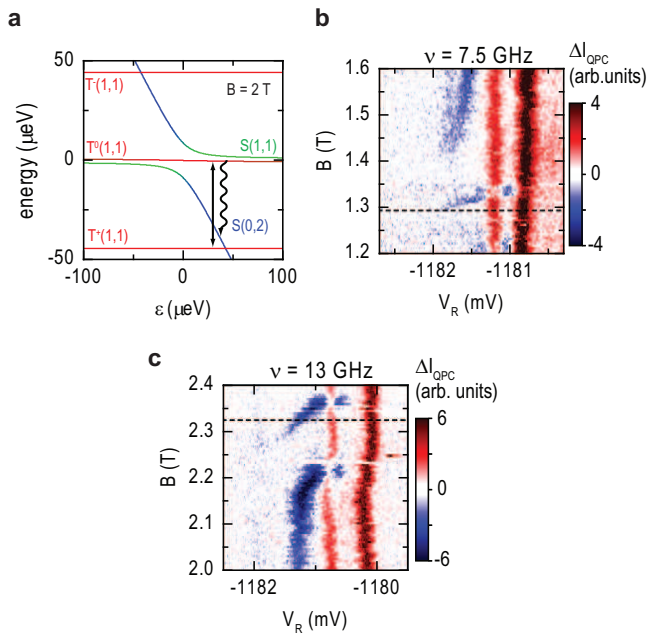}
\caption{\label{fig5}%
\textbf{Triplet resonance in the PAT spectra.} \textbf{a}, Energy eigenstates of the two-electron spin states plotted with the same color code as in Fig. 2c of the main article. Microwaves are resonant to the $T^+(1,1)$/$T^0(1,1)$ transition. Spontaneous relaxation from $T^0(1,1)$ to the metastable $S(0,2)$ enables the detection via $\Delta I_{QPC}$. \textbf{b}, The PAT spectrum as measured with a microwave frequency of 7.5~GHz exhibits a nearly horizontal blue feature as visible in Fig. 2f. Also at this microwave frequency the line starts from the magnetic field, at which we expect the electron spin resonance condition to be fulfilled (black dashed line) $h\nu=g \mu (B+b_0)$. \textbf{c}, PAT spectrum recorded at 13~GHz with corresponding line indicating electron spin resonance.}
\end{figure}

In Fig. 2f of the main article, a PAT feature is observed that is due to a transition from the $T^+(1,1)$ ground state to the $T^0(1,1)$ excited state. The $T^0(1,1)$ state can relax via spontaneous phonon emission to the singlet bonding state, a superposition of $S(1,1)$ and $S(0,2)$. This state is metastable, since the spontaneous phonon relaxation is suppressed by the small energy difference to the $T^+(1,1)$ ground state (see Fig. S2a), which makes the transition detectable by $\Delta I_{QPC}$. A peculiarity of the $T^+(1,1)$ to $T^0(1,1)$ resonance is its slope in the PAT spectrum, which might be a result from a gradient magnetic field along the magnetic field direction as discussed in the main article. Note that in the same detuning range, we observe also direct PAT transitions from the $T^+(1,1)$ ground state to the singlet anti-bonding state and singlet bonding state at lower and higher magnetic fields, respectively.

Here we investigate the position of the $T^+(1,1)$ to $T^0(1,1)$ resonance, as a function of the microwave frequency $\nu$. The Figs. S2b and S2c show raw PAT spectra (without any post-processing step applied as explained above) recorded with $\nu=7.5$~GHz and $\nu=13$~GHz, respectively. The dashed lines mark the magnetic field, at which the electron spin resonance condition $h \nu = g \mu_B (B + b_0)$ is fulfilled. Here we use the absolute electronic g-factor $|g|=0.382$ and longitudinal magnetic field offset $b_0=109$~mT as determined by the Figs. 3a and 3b of the main article. Alternatively, we might use the $T^+(1,1)$ to $T^0(1,1)$ resonance feature to determine $|g|$. If we use the center magnetic field of this feature as the resonant field, we calculate $|g|=0.384 \pm 0.005$ from the microwave frequency dependence in good agreement with the $|g|=0.382$ found in Fig. 3a in the main article.

\section{The $\Delta m=0$ PAT transition}
In the main article, we discuss whether the red ($\Delta m=0, \Delta n=1$) PAT resonance is dominantly due to a transition from the $S(0,2)$ ground state to the $S(1,1)$ state or to the $T^0(1,1)$ state. It is difficult to spectroscopically resolve these transitions, since they differ only by the exchange energy $J(\varepsilon)$. In Fig 3a of the main article, we use $\Delta \varepsilon^\pm(B)$, the difference in detuning between the red $\Delta m=\pm 1$ and the $\Delta m=0$ lines, to assign the PAT resonance. Here, we support this argument by calculating the expected $\Delta \varepsilon^\pm(B)$ functions for both extreme scenarios: a pure $S(0,2)$/$S(1,1)$ and a pure $S(0,2)$/$T^0(1,1)$ transition. For the calculation, we use the determined $|g|=0.382$, $t_c=8.7$~$\mu$eV and $b_0=109$~mT values. The result is displayed in Fig. S3a for a microwave frequency $\nu=20$~GHz. Obviously, the scenario of a pure singlet transition results in $\Delta \varepsilon^+(B)>\Delta \varepsilon^-(B)$, whereas $\Delta \varepsilon^+(B) = \Delta \varepsilon^-(B)$ is found for a transition to the triplet state. In the experiment, we clearly observe $\Delta \varepsilon^+(B)>\Delta \varepsilon^-(B)$ and therefore the $\Delta m=0$ transition is dominantly a singlet transition.

\begin{figure}
\includegraphics[width=8.5cm]{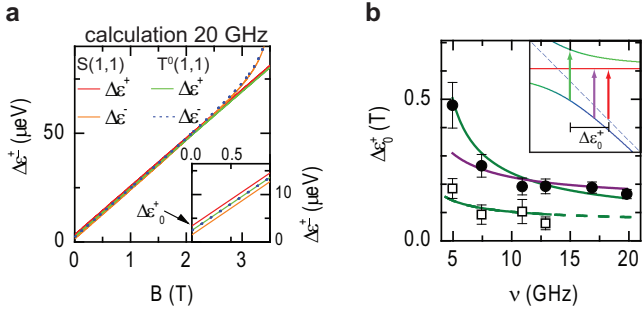}
\caption{\label{fig3}%
\textbf{Analysis of the $\Delta m=0$ PAT resonance.} \textbf{a}, $\Delta \varepsilon^+$ and $\Delta \varepsilon^-$ calculated for the parameters $|g|=0.382$, $t_c=8.7$~$\mu$eV, $b_0=109$~mT, $\nu=20$~GHz and two possible scenarios: The $\Delta m=0$ line is resonant between the $S(0,2)$ ground state $S(1,1)$ (red and orange line) and $T^0(1,1)$ (green and blue line). Zoom-in for low-magnetic fields in the inset.\textbf{b}, The detuning offset $\Delta \varepsilon_0^+$ between the transition $\Delta m= +1$ for $B \rightarrow 0$~mT (red arrow in the inset) and the $\Delta m= 0$ transitions from $S(0,2)$ to $S(1,1)$ (green arrow) and to $T^0(1,1)$ (violet arrow), respectively, is plotted as a function of the microwave frequency $\nu$. The error bars are determined from the linear extrapolation $B \rightarrow 0$~mT. Shown are offsets measured at $t_c = 8.7$~$\mu$eV (filled circles) and at $t_c \approx 4$~$\mu$eV (open rectangles). The green and violet lines are least-squares fits to the measured offsets assuming a pure $S(0,2)$ to $S(1,1)$ and to $T^0(1,1)$ transition, respectively.
}
\end{figure}

In both scenarios, the $\Delta \varepsilon^-(B)$ increases non-linearly towards high $B$, since the resonance $\Delta m=-1$ becomes sensitive to the curved singlet anti-crossing at $E_z \approx h\nu-t_c$, where $E_z$ is the Zeeman energy. This curvature is not observed in the experiment as the $\Delta m=-1$ transition fades out at $B \approx 1.5$~T. In both scenarios, the $\Delta \varepsilon^+(B)$ functions are linear, which holds true only if $E_z \ll t_c-h\nu$. The choice of a high $\nu$ and an appropriate $B$-range allows to extract $|g|$ from $\Delta \varepsilon^+(B)$ by a simple linear least-squares fit as demonstrated in the main article.

As a final step, we analyze the splitting of the $\Delta \varepsilon^+(B)$ and $\Delta \varepsilon^-(B)$ function quantitatively. Their offsets depend upon the exchange energy $J(\varepsilon,t_c)$, which we can experimentally vary by $t_c$ or indirectly by $\nu$. For $h \nu \rightarrow 2 t_c$, the detuning position of the $\Delta m=0$ resonance becomes strongly affected by $J$. As shown in the inset of Fig. S3b, the $S(0,2)$/$S(1,1)$ transition (green arrow) shifts more towards negative detuning $\varepsilon$ than the potential $S(0,2)$/$T^0(1,1)$ transition (violet arrow). We measure $\Delta \varepsilon^+(B)$ for various $\nu$ and determine the $\Delta \varepsilon^+_0=\Delta \varepsilon^+(B \rightarrow 0)$ by a linear fit at a sufficiently high magnetic field range. This procedure turned out to be impractical with $\Delta \varepsilon^-_0(B)$, because the $\Delta m=-1$ line fades out at high magnetic fields. Note that the linear extrapolation of the $S(0,2)/T^+(1,1)$ transition to zero field (red arrow) exhibits a different detuning position than the $S(0,2)$/$T^0(1,1)$ transition (violet arrow), because the linear extrapolation follows the dashed blue line in the inset of Fig. S3b, i.e. the linear extrapolated detuning position is not affected by the hybridization of the singlets. The detuning position of the $S(0,2)$/$T^0(1,1)$ PAT transition, however, is affected by the singlet hybridization, since it lowers the energy of the initial state $S(0,2)$. In summary, the value of $\Delta \varepsilon^+_0$ is always larger then zero, but also depends upon the nature of the $\Delta m=0$ PAT resonance.

In the inset of Fig. S3b, $\Delta \varepsilon^+_0$ for the $S(0,2)$/$S(1,1)$ transition is drawn. Obviously, $\Delta \varepsilon^+_0$ is considerably smaller, if the $S(0,2)$/$T^0(1,1)$ PAT resonance dominates over the $S(0,2)$/$S(1,1)$ resonance. The analysis of $\Delta \varepsilon^+_0$ is complicated by the remanence of the $\mu$magnet, which is not exactly known, but should be smaller than the fully magnetized field of $b_0 = 109$~mT. Due to the remanence, the linear extrapolation of the $S(0,2)$/$T^+(1,1)$ transition towards zero external magnetic field, leaves an additional offset on $\Delta \varepsilon^+_0$. This offset, however, is independent from $\nu$ and $t_c$. In Fig. S3b, the extrapolated $\Delta \varepsilon^+_0$ values are plotted as a function of the microwave frequency $\nu$ for two tunnel couplings (filled circles and open squares). Fitting the filled circles with the well-known $t_c = 8.7$~$\mu$eV, we determine a reasonable fit by assuming the $\Delta m=0$ PAT resonance to be purely singlet (green line). The fit with a potential $S(0,2)$/$T^0(1,1)$ transition (violet line) fails at $\nu=5$~GHz. The only fit parameter used here is the remanence of the $\mu$magnet, which was found to be $\approx 70$~mT for the fit function assuming a $S(0,2)$/$S(1,1)$ transition, and $\approx 140$~mT assuming a $S(0,2)$/$T^0(1,1)$ transition. The latter is very unlikely, since only a maximum magnetization of $b_0=109$~mT was found at an external magnetic field of 2~T.

As a final check, we take $\Delta \varepsilon^+_0$ values into account, which were determined when the DD was tuned to a smaller $t_c = 4$~$\mu$eV (open squares). These data points cannot be fitted by the fit function that assumes a $S(0,2)$/$T^0(1,1)$ transition at all, since the $\Delta \varepsilon^+_0/g \mu_B$ values observed are already smaller than the remanence of $\approx 140$~mT, which would stay valid for the altered tunnel coupling. Obviously, this leads to a contradiction, since all $\Delta \varepsilon^+_0$ would become negative after subtracting the remanence. Only the assumption of a purely singlet $\Delta m=0$ PAT transition in combination with the smaller remanence of $\approx 70$~mT, as fitted above, allows reasonable fitting. Our conclusion from the main article is therefore further supported.

\section{Spin flip-tunneling mechanism}

As noticed in the main text, there is a direct matrix element between the $S(0,2)$ and the $(1,1)$ triplet states.
This can occur due to the nuclear spins in the barrier
between the dots and due to spin-orbit (SO) interaction.  We
look at a toy model to examine the relative importance of these two
processes.  Specifically, we consider the hopping matrix element for a single electron with spin $\vec{S}$
between two wavefunctions associated with an electron on the left
($\ket{L}$) and on the right ($\ket{R}$) via the perturbation:
\begin{eqnarray}
V & = & \frac{\hbar}{m^* \lambda_{SO}} [-\alpha (S_{\tilde{z}} p_{\tilde{y}} - S_{\tilde{y}} p_{\tilde{z}}) + \beta (p_{\tilde{y}} S_{\tilde{y}} + p_{\tilde{z}} S_{\tilde{z}})] + \nonumber \\
& & + A v_0 \sum_j \delta(r - r_j) \vec{I}^j \cdot \vec{S}
\label{SO}
\end{eqnarray}
where we have absorbed the Rashba ($\alpha$) and Dresselhaus ($\beta$)
terms into a single spin-orbit interaction with a characteristic
spin-orbit length $\lambda_{SO} \sim 10\ \mu$m.  We recall that $m^*$ is the effective electron mass, $A
\approx$~$100 \mu$eV, $v_0$ is the unit cell volume and $I^j$ is the nuclear spin at $\vec{r}_j$.

We now wish to estimate the spin-flip tunneling for the single electron case, given by averaging
over the orbital dipole:
\begin{equation}
\bra{R} V \ket{L}
\end{equation}
This can be evaluated explicitly for $\ket{L(R)} = \exp(-(z \pm a/2)^2/4\sigma^2)/(2 \pi \sigma^2)^{1/4} \phi(x,y)$ where
$z$ is the inter-dot axis (at an angle $\theta$ with the
axis $\tilde{z}$ from the spin-orbit interaction in Eq. \ref{SO}) and $\phi(x,y)$ is the
transverse-longitudinal wavefunction. As tunneling occurs only along the $z$-axis, matrix elements with $p_y$ are zero. We find two tunneling matrix elements:
\begin{eqnarray}
t_{SO} & = & \frac{\hbar^2}{m^* \sigma^2} \frac{a}{4 \lambda_{SO}} e^{-a^2/8\sigma^2}
\vec{n} \cdot \vec{S} \\
t_{nuc} & = & g \mu_B \vec{B}_{\rm nuc,f} \cdot \vec{S} \\
\vec{n} & = & - \cos(\theta) [(\alpha - \beta) \cos(\theta) +
(\alpha+\beta) \sin(\theta) ] \hat{z} - \nonumber \\
& & \sin(\theta) [(\beta - \alpha) \sin(\theta) + (\alpha + \beta) \cos(\theta)] \hat{y} \\
g \mu_B \vec{B}_{\rm nuc,f} & = & A v_0 \sum_j |\psi_L(r_j)||\psi_R(r_j)| \vec{I}^j
\end{eqnarray}
We remark that the rms value for $\vec{B}_{\rm nuc,f}$ is given by
\begin{eqnarray}
  g \mu_B \sqrt{|\vec{B}_{\rm nuc,f}|^2} & = & A \sqrt{ v_0^2 \sum_j
    |\psi_L|^2 |\psi_R|^2 I(I+1)} \nonumber \\
    & \approx & e^{-a^2 / 8 \sigma^2} A
  \sqrt{v_0^2 \sum_j |\psi_L|^4}
\end{eqnarray}
That is, it is the rms value for a single dot, $A/\sqrt{N}$, multiplied by $\exp(-a^2
/ 8 \sigma^2)$.  Also, the size of the single-particle wavefunction,
$\sigma$, is related to the orbital energy scale of a single dot by
$\Delta \approx \frac{\hbar^2}{m_* \sigma^2}$.  Thus, the relative
strength of the two tunneling terms (including spin flip) is
\begin{equation}
  \frac{|t_{SO}|}{ |t_{nuc}|} = \frac{\Delta}{A / \sqrt{N}}
  \frac{3|\vec{n} \times \vec{B}_{\rm ext}|}{2|B_{\rm ext}|} \frac{a}{4 \lambda_{SO}}
\end{equation}
where $N$ is the number of spins in a single quantum dot.

We explore briefly how this ratio varies with dot size $\sigma$ and
spacing $a$.  Specifically, $\Delta \sqrt{N} \propto \sigma^{-1}$,
so a larger dot reduces the strength of spin-orbit tunneling compared
to hyperfine-assisted tunneling.  On the other hand, increasing the distance
$a$ increases the relative strength of spin-orbit tunneling to
hyperfine-assisted tunneling. Setting in $A=100$~$\mu$eV, $N=4 \times 10^6$, $a=75$~nm, $\Delta=1000$~$\mu$eV and $\lambda_{SO} = 10$~$\mu$m, we calculate the ratio of the matrix elements $\frac{|t_{SO}|}{ |t_{nuc}|}$ to be $\approx 60$. We remark that in external field parallel to $\vec{n}$ prevents any SO spin-charge flips. All spin charge flips occur only via $t_{nuc}$.

In the bases ($T^-(1,1)$, $|\downarrow \uparrow \rangle$, $|\uparrow \downarrow \rangle$, $T^+(1,1)$, $S^-(0,2)$) the two electron Hamiltonian that ignores the small $t_{nuc}$ term has the form

\begin{widetext}
\begin{equation}
\left( \begin{array}{ccccc}
B_z & \frac{\Delta B_x^\bot - i \Delta B_y^\bot}{2} & -\frac{\Delta B_x^\bot - i \Delta B_y^\bot}{2} & 0 & -t_{SO,y}/\sqrt{2} \\
\frac{\Delta B_x^\bot + i \Delta B_y^\bot}{2} & -\Delta B^\| & 0 & -\frac{\Delta B_x^\bot - i \Delta B_y^\bot}{2} & (-i t_{SO,z} - t_c)/\sqrt{2} \\
-\frac{\Delta B_x^\bot + i \Delta B_y^\bot}{2} & 0 & \Delta B^\| & \frac{\Delta B_x^\bot - i \Delta B_y^\bot}{2} & (-i t_{SO,z} + t_c)/\sqrt{2} \\
0 & -\frac{\Delta B_x^\bot + i \Delta B_y^\bot}{2} & \frac{\Delta B_x^\bot + i \Delta B_y^\bot}{2} & -B_z & -t_{SO,y} \\
-t_{SO,y}/\sqrt{2}  &  (i t_{SO,z} - t_c)/\sqrt{2} &  (i t_{SO,z} + t_c)/\sqrt{2}  & -t_{SO,y}/\sqrt{2}  & - \varepsilon
\end{array} \right)
\end{equation}
\end{widetext}

where, e.g., the Larmor precession frequency of an electron spin in the left dot is $B_z - \Delta B^\|$.

\section{Simulations of the PAT spectra - relaxation}

\begin{figure}
\includegraphics[width=5 cm]{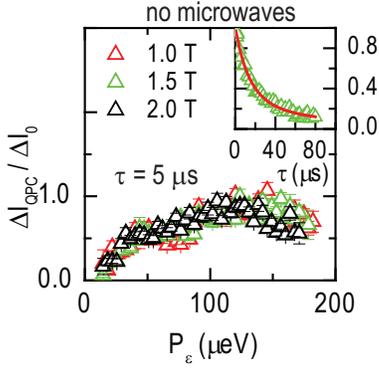}
\caption{\label{fig7}%
\textbf{Spontaneous relaxation rate.} Normalized QPC current averaged over $\tau = 5 \mu$s immediately after full mixing at the $ST^+$ anticrossing for various magnetic fields at zero microwave power. The spontaneous spin relaxation after mixing is measured at a distance $P_\varepsilon$ to the $ST^+$ mixing point (see scheme in Fig. 2c). The dependence on the averaged time interval $\tau$ is displayed in the inset for $B=1.5$~T and $P_\varepsilon=2$~mV together with a least-squares fit (red solid line). }
\end{figure}

The simulated spectra in Figs. 2c,d of the main article include the effect of the phonon-mediated relaxation. In addition to the explanations of the simulations in appendix A, we continue here on the coupling of the electron spin to the phonon bath.
We thereby neglect deformation phonons as the energy scales examined in the experiment (7-22 GHz) are much smaller than the characteristic frequency scale of a phonon on the length scale of the dot $c_{ph}/ l_{dot} \sim 60-120$~GHz.  To determine the coupling, we take as an ansatz for the electronic wavefunctions the Fock-Darwin states, given by Gaussians, and calculate the coupling after orthogonalizing the states with the perturbation $V_{ph} = \sum_{e,k} f(k_z) \sqrt{\frac{\hbar}{\rho \omega_{e,k}}} e^{i \vec{k} \cdot \vec{r}} \beta_e (a_{e,k} - a_{e,-k}^\dag)$ \cite{Kittel}, where $f(k_z) \approx 1$ for the energy scales we are working with.  We then use Fermi's golden rule to calculate excitation and relaxation from thermal and spontaneous emission of phonons. Finally, we numerically solve the superoperator for the steady state and compare the expectation value of $\ketbrad{S(0,2)}$ with and without the excitation $\Omega$, mimicking the effect of the lock-in detection.

\section{Measurement of spontaneous relaxation}

\begin{figure*}
\includegraphics[width=12.0cm]{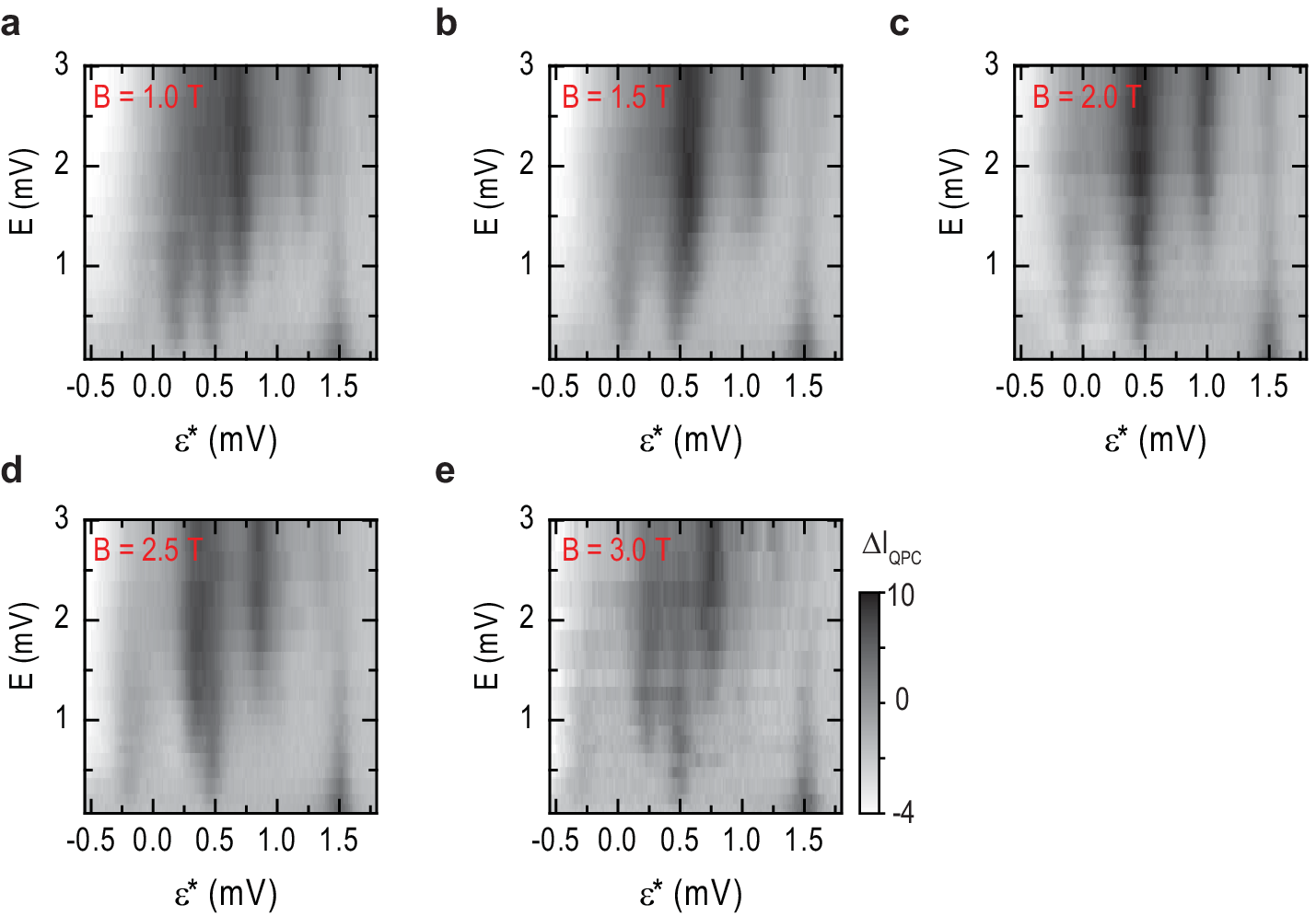}
\caption{\label{fig7}%
\textbf{Power-dependence of the PAT spectra for different magnetic fields.} \textbf{a-e}, Microwave induced change of the QPC current as a function of the double dot detuning $\varepsilon$ and the microwave amplitude $E$ at the sample for 11~GHz. From \textbf{a} to \textbf{e}, the external magnetic field is increased by $500$~mT. The detuning axis is calibrated so that $\varepsilon^*=0$~mV is equal to the detuning position at the $S(1,1)$-$T^+(1,1)$ anti-crossing (see Fig. 1c for the calibration method). The pulse amplitude of the reference line is $P_\varepsilon=1.5$~mV.}
\end{figure*}

In the main article, we state that the spontaneous relaxation rate from $T^+(1,1)$ to $S(0,2)$ is found to be magnetic field independent. Relevant for the PAT process is the spontaneous relaxation at a fixed energy difference between the $T^+(1,1)$ and $S(0,2)$ state, which is set by the photon energy. Thus, when changing the magnetic field, the detuning $\varepsilon$ has to be changed accordingly. The measurement of the spontaneous relaxation rate is done as follows: Starting from $S(0,2)$, we populate the $T^+(1,1)$ by $50$~\% via a $200$~ns detuning pulse \cite{Barthel09} with amplitude $P_\varepsilon$ in the absence of microwaves, and monitor the decay back to $S(0,2)$. The relaxation rate $\Gamma_s$ can be extracted from the time averaged lock-in signal $\Delta I_{QPC}(\tau) = \frac{\Delta I_0}{\tau} \int_{0}^\tau \exp\left(-t \Gamma_s \right) dt$, where $\tau$ is the time spent in Pauli blockade between the pulses. $\Delta I_0=\Delta I_{QPC}(0)$, which is independent from $B$, is extracted from the fit in the inset of Fig. S4. In order to cover various values of the detuning and the magnetic field, we next fix $\tau = 5$~$\mu$s and record $\Delta I_{QPC}$ as a function of $P_\varepsilon$ for three different magnetic fields. We observe that $\Delta I_{QPC}(5~\mu$s$)$ and thus $\Gamma_s$ are essentially independent of $B$ (Fig. S4). This holds true for all $P_\varepsilon$ and hence for all $T^+(1,1)$-$S(0,2)$ energy splittings. Note that regardless of $B$, this energy splitting is given by $P_\varepsilon$ alone. This reflects exactly the situation in the PAT experiment, for which the microwave frequency alone sets the energy splitting and thus also the required phonon energy for the spontaneous relaxation process.

\section{Power dependence}
In Fig. 4a of the main article, the power dependence of the PAT sidebands is shown for two magnetic fields. In Fig. S5a-e, $\nu = 11$~GHz-spectra measured with a series of magnetic field values are plotted, to ease keeping track of the resonances as they shift in detuning with the external magnetic field. Here, the spectra are plotted such that $\varepsilon^*=0$ is the $ST^+$ anti-crossing for all $B$ (compare Fig. S1c). The reference peaks due to pulsing to the $ST^+$ anti-crossing are all aligned at $\varepsilon^*=P_\varepsilon=1.5$~mV, which is the pulse amplitude. The amplitude of the microwaves $E$ is estimated from the attenuation of the high-frequency circuit at room temperature.

PAT sidebands at larger detuning appear when $E$ is increased as expected for PAT. Oscillation of the PAT amplitude as a function of $E$ is hardly visible, because we cannot reach sufficiently high $E$ and because the PAT lines become broadened as a function of power. One sideband emerges at high $\varepsilon^*\approx 0.5$~mV already at low $E$. This resonance stays at a constant $\varepsilon^*$ for all $B$ and is therefore a transition from $S(0,2)$ to $T^+(1,1)$ ($\Delta m=1$). As $B$ is increased, the other transitions move towards lower $\varepsilon^*$ while keeping the distance along $\varepsilon^*$. They are due to PAT transition with $\Delta m =0$. At $B \approx 2$~T, the $\Delta m=1$ PAT transition overlaps with the 2-photon $\Delta m=0$ PAT transition. This is expected, since $\nu=11$~GHz equals the Zeeman energy at this magnetic field in our double dot.

\begin{acknowledgments}
 We gratefully acknowledge discussions with S. M. Frolov, M. Laforest, D. Loss, Yu. V. Nazarov, K. C. Nowack, M. Shafiei and thank H. Keijzers for help with the sample fabrication and R. Schouten, A. van der Enden and R. G. Roeleveld for technical support. This work is supported by the `Stichting voor Fundamenteel Onderzoek der Materie (FOM)' and a Starting Investigator grant of the `European Research Council (ERC)'.

 L.R.S, F.R.B., V.C. and T.M. performed the experiment, W.W. grew the heterostructure, T.M. fabricated the sample, L.R.S., J.D., J.M.T and L.M.K.V. developed the theory, J.M.T. did the simulations, all authors contributed to the interpretation of the data and commented on the manuscript, and L.R.S., J.D., J.M.T. and L.M.K.V. wrote the manuscript.
\end{acknowledgments}

\end{document}